\documentclass[class=preprint]{nature}
\usepackage{graphicx}
\graphicspath{{../../Figures/}}
\usepackage[usenames,dvipsnames]{color}
\usepackage[colorlinks=true,citecolor=blue,linkcolor=blue,urlcolor=blue]%
{hyperref}
\usepackage[a4paper]{geometry}
\usepackage{fixltx2e}
\usepackage{upgreek}
\usepackage{amsmath}
\usepackage{amssymb}
\usepackage{amsthm}
\usepackage{mathrsfs}
\usepackage{amsbsy}
\usepackage{amstext}
\usepackage{xcolor}
\usepackage{float}
\usepackage{multirow}
\usepackage{wasysym}
\usepackage{mathtools}
\usepackage{subfigure}
\usepackage{mhchem}
\usepackage{amsfonts}
\usepackage{dcolumn}
\usepackage{placeins}
\usepackage{float}
\usepackage{caption}
\usepackage[right]{lineno}
\usepackage{fancyhdr}
\usepackage{bbold,bm}
\usepackage{pdfpages}
\usepackage[utf8]{inputenc}
\usepackage[T1]{fontenc}%
\usepackage{setspace}
\providecommand{\U}[1]{\protect\rule{.1in}{.1in}}
\DeclareCaptionLabelSeparator{vline}{$\boldsymbol{:}$}
\title{Terahertz field-induced nonlinear coupling of two magnon modes in an antiferromagnet}
\author{Zhuquan Zhang$^{1\dagger}$, Frank Y. Gao$^{2\dagger}$, Jonathan B. Curtis$^3$, Zi-Jie Liu$^1$, Yu-Che Chien$^1$, Alexander von Hoegen$^4$, Man Tou Wong$^1$, Takayuki Kurihara$^5$, Tohru Suemoto$^5$, Prineha Narang$^3$,  Edoardo Baldini$^{2*}$, and Keith A. Nelson$^{1*}$ \\
	\normalsize{$^1$Department of Chemistry, Massachusetts Institute of Technology, Cambridge, Massachusetts, USA, 02139 }\\
	\normalsize{$^2$Department of Physics, The University of Texas at Austin, Austin, Texas, USA, 78712 } \\
	\normalsize{$^3$College of Letters and Science, University of California, Los Angeles, California, USA, 90095}\\
	\normalsize{$^4$Department of Physics, Massachusetts Institute of Technology, Cambridge, Massachusetts, USA, 02139 }\\ 	
	\normalsize{$^5$Institute for Solid State Physics, The University of Tokyo, Kashiwa, Chiba, Japan, 277-8581} \\ 
	\normalsize{$^{*}$E-mail: kanelson@mit.edu, edoardo.baldini@austin.utexas.edu} \\
	\normalsize{$^\dagger$These authors contributed equally to this work} \\
	\\
}
\begin{document}

\maketitle

\section*{Abstract}
Magnons are quantized collective spin-wave excitations in magnetically ordered materials. Revealing their interactions among these collective modes is crucial for the understanding of fundamental many-body effects in such systems and the development of high-speed information transport and processing devices based on them. Nevertheless, identifying couplings between individual magnon modes remains a long-standing challenge. Here, we demonstrate spectroscopic fingerprints of anharmonic coupling between distinct magnon modes in an antiferromagnet, as evidenced by coherent photon emission at the sum and difference frequencies of the two modes. This discovery is enabled by driving two magnon modes coherently with a pair of tailored terahertz fields and then disentangling a mixture of nonlinear responses with different origins. Our approach provides a route for generating nonlinear magnon-magnon mixing.
\newpage

\section*{Main Text}
 
Understanding and controlling the interactions between quasiparticles lies at the core of condensed matter research.\cite{turner2010coherent, forst_nonlinear_2011, kozina2019terahertz, disa2020polarizing ,mashkovich2021terahertz, bae2022exciton} In magnetically ordered systems, the collective excitations of spin precessions are spin waves – the quanta of which are magnons. The coupling between multiple magnons not only leads to exotic phenomena in quantum magnetism, such as magnon bound states\cite{bethe1931theorie, wortis1963bound}, frustrated interactions\cite{bai2021hybridized}, and magnon Bose-Einstein condensation\cite{demokritov2006bose,borisenko2020direct}, it also lays the foundation for various applications leveraging coherent nonlinear magnonics\cite{pirro2021advances, li2022perspective}. For the latter, the nonlinear couplings between magnetic excitations could enable fundamentally new ways of manipulating spin waves that are beyond what can be achieved by linear superpositions, and hold promise for developing all-magnonic signal processing devices. Such protocols have been successfully implemented in the microwave frequency regimes for ferromagnets and ferrimagnets, in which multiple spin wave signals generated at different wavevectors for the same magnon branch interact with each other in a nonlinear fashion, thereby manifesting as nonlinear spin-wave mixings\cite{wang2021magnonic,carmiggelt2023broadband}. Pertaining to antiferromagnets, which exhibit more intricate magnon modes in the terahertz (THz) frequency range, distinct magnon modes can potentially intermix nonlinearly if both modes are driven out-of-equilibrium into the anharmonic coupling regime. This subsequently gives rise to Manley–Rowe dynamics\cite{haus1991coupled,boyd2008nonlinear} of magnon coherences, bearing resemblance to difference- and sum-frequency generations (DFG and SFG) observed in nonlinear optics. As such, the signatures of these coherent couplings provide exclusive evidences of magnon-magnon interactions that are not captured by most steady-state measurements, as they usually appear only under special circumstances, such as when an external field brings magnon frequencies into near-degeneracy\cite{macneill2019gigahertz, makihara2021ultrastrong,lee2023nonlinear}. In contrast to those multi-magnon interactions observed via incoherent scattering,\cite{songvilay_anharmonic_2018},  these coherent couplings offer a dynamic tuning knob to control correlated magnonic states that are not present in thermodynamic equilibrium.

However, coherent magnon-magnon mode mixings are usually hidden in conventional spectroscopies dominated by linear magnon excitations\cite{kampfrath2011coherent}, and the existence of these coherent couplings as well as their exact excitation pathways remain hitherto yet to be established. Here, we explore an uncharted territory in nonlinear coupled magnonics by measuring the coherence and interactions between two distinct magnon modes in the canted antiferromagnet YFeO$_3$ at room temperature. The dynamics of interacting magnons are imprinted on the nonlinear electrodynamic response that emerges upon the simultaneous driving of modes through a pair of THz pulses. By transforming the time-domain response into a two-dimensional (2D) frequency-frequency map, we uncover the nonlinear mixing of anharmonically coupled magnon modes. This discovery was made possible by a state-of-the-art 2D THz polarimetry technique utilizing single-shot detection of coherent THz signals\cite{teo2015invited,gao2022high}. This method allows us to rapidly collect over a hundred 2D THz spectra with various crystalline orientations with respect to the incident THz field polarization, enabling us to observe anisotropic responses that arise from nonlinear magnon-magnon mixing. 

Figure 1a depicts the crystal structure and spin configurations of YFeO$_3$. As a model antiferromagnetic insulator, YFeO$_3$ crystallizes in an orthorhombically distorted perovskite structure, with the two nearest neighboring Fe$^{3+}$ ions ordered nearly antiparallel to the $a$ axis. However, due to the Dzyaloshinskii–Moriya interaction\cite{dzyaloshinsky1958thermodynamic, moriya1960anisotropic}, the spins are slightly canted, leading to a net magnetization along the $c$ axis. This peculiar magnetic structure allows for two primary cooperative motions of sublattice spins, corresponding to two distinct magnon modes\cite{lu2017coherent, yamaguchi2010coherent}, i.e., the quasi-ferromagnetic (qFM) mode, which corresponds to a precession of the magnetization orientation, and the quasi-antiferromagnetic (qAFM) mode, which corresponds to an oscillation of the magnetization amplitude. The real-space spin precessions of both are displayed in Fig. 1a. 

As a first step, we study how the magnon modes in YFeO$_3$ respond to a single linearly polarized THz pulse. As shown in Fig. 1b, a single-cycle THz transient is focused on a (010)-cut YFeO$_3$ sample, and the radiated free induction decay (FID) signals are tracked in the time-domain by our single-shot measurement technique\cite{teo2015invited, gao2022high}, which yields the field amplitudes at 500 different detection times $t$ given by the times at which each of 500 optical readout pulses overlaps with the THz signal field inside an electro-optic crystal. By rotating the sample, we can selectively drive either magnon mode or both modes, depending on the relative orientations of the THz field polarization and the crystallographic axes. Figure 1c shows the time-domain FID signals when the THz magnetic field is oriented along the \textit{a}, \textit{c}, or \textit{ac} (i.e., 45$^\circ$ to \textit{a} and \textit{c} axes) directions. In the \textit{ac} orientation, the crystal birefringence leads to a temporal walk-off of the THz field components along the different crystallographic axes\cite{yamaguchi2010coherent,jin2013single}, resulting in a signal with two peaks (see Supplementary Note 1). The FID signals show two types of oscillations which are assigned to the previously reported qFM ($\Omega_{qFM} = 0.30$ THz) and qAFM ($\Omega_{qAFM} = 0.53$ THz) magnon modes\cite{yamaguchi2010coherent} as evidenced by the Fourier transforms in Fig. 1d.

To further reveal the anisotropic responses of the driven magnon modes, we use THz polarimetry to follow the azimuthal dependence of each signal. We perform this measurement by rotating the sample in 5$^\circ$ intervals, completing a full 360$^\circ$ sweep and measuring the resulting FID signal polarizations both parallel ($\mathbf{H}_{det} \parallel \mathbf{H}_{THz}$) and perpendicular ($\mathbf{H}_{det} \perp \mathbf{H}_{THz}$) to the incident THz field polarization, controlled with a wire grid polarizer. We extract the amplitudes of each magnon mode from the Fourier spectra of the polarimetry signals and plot them as a function of the azimuthal angle, $\theta$, relative to the $a$ crystallographic axis, as shown in Figs. 1e and 1f. From these data, it is evident that only a single magnon mode is excited when the THz magnetic field is polarized along either crystallographic axis (Fig. 1e), in line with the expected Zeeman interaction-induced out-of-phase or in-phase rotations of two sublattice spins. However, when the THz polarization is off-axis, the field interaction has components that simultaneously induce excitation of both modes (Fig. 1f). The azimuthal dependence agrees with excitation of magnon modes via the Zeeman interaction with the magnetic component of the THz field.

To uncover the inherent nonlinear couplings between the two distinct magnon modes, we add another THz pulse of the same polarization with variable time delay $\tau$ before the existing one. For each $\tau$, we record the time-dependent coherent THz signal fields originating from either and both THz pulses using the single-shot measurement method\cite{zhang2022nonlinear,gao2022high}. We can then isolate the weak nonlinear response, which is otherwise hidden in the FID signal, by calculating the difference between these signals, see Eq. (1) in the Methods section and previous reports\cite{lu2017coherent}. 2D Fourier transforms of these nonlinear signals $S(\tau,t)$ produce frequency-frequency correlation maps $S(\nu,f)$ for three different incident THz magnetic field orientations (Fig. 2d: $a$ axis, 2e: $c$ axis and 2f: $ac$ bisector). THz magnetic fields orientated along the \textit{a} and \textit{c} axes show only excitation of the qFM (i.e. I) and qAFM (i.e. II) magnon modes respectively. In this case, we observe third-order nonlinear optical responses arising from degenerate four-spin-wave mixing including non-rephasing (NR, $[\Omega,\Omega]$), rephasing (R, i.e. photon echo, $[\Omega,-\Omega]$), pump-probe (PP, $[\Omega,0]$) and two-quantum (2Q, $[\Omega,2\Omega]$) signals, in addition to second-order nonlinear optical responses including THz rectification (TR, $[0,\Omega]$), and second-harmonic generation (SHG, $[2\Omega,\Omega]$) signals. These observations are consistent with previous measurements which only involved excitation of individual magnon modes in YFeO$_3$\cite{lu2017coherent} -- see Supplementary Note 2 for a detailed description of the peak assignments and their origin. 2D THz spectra measured with the THz magnetic field oriented  along the \textit{ac} bisector show excitations of both FM and AFM magnon modes as evidenced by nonlinear peaks corresponding to individual excitations of either mode. In addition to these signals, we also observe a set of four emergent peaks corresponding to the excitation of both qFM and qAFM modes, in either temporal order. A pair of these signals (SFG I and SFG II) appear at the sum-frequency of the two magnon modes $\Omega_{qFM}+\Omega_{qAFM}$ ($0.83$ THz), while the other signals (DFG I and DFG II) appear at the difference frequency $\Omega_{qAFM}-\Omega_{qFM}$ (0.23 THz) along the detection axis in the 2D spectrum. All these spectra are reproduced in spin dynamics simulations employing the Landau-Lifshitz-Gilbert (LLG) equation, see Extended Data Fig. 1. Our labeling of the SFG and DFG peaks with mode I or II indicates that as the time $\tau$ between THz pulses was varied, the signal amplitude showed oscillations at frequency $\Omega_{qFM}$ or $\Omega_{qAFM}$ respectively, i.e. at the excitation frequency in the 2D spectrum. Unlike previous coherent 2D optical spectroscopy studies, where the anharmonicity of system is inferred in the dispersive features of diagonal peaks\cite{khalil2001signatures,singh2016polarization}, magnon-magnon signals occurs as the result of coherent anharmonic coupling between the two distinct magnon modes.

To elucidate the genesis of these peaks, we perform both field-dependent and THz polarimetry measurements of the 2D spectra. Figures 3a and 3d show the resulting field dependences of the peak amplitudes of the SFG and DFG signals. In all cases, the amplitudes of these signals scale quadratically with the incident THz magnetic field, consistent with their assignment to magnetic $\chi_m^{(2)}$ processes. For the 2D THz polarimetry measurements, the sample is rotated as in our linear polarimetry measurements and parallel and perpendicular-polarized 2D THz spectra are collected for each angle. Polar plots depicting the peak amplitudes of SFG and DFG signals as a function of the incident THz magnetic field orientation are shown in Figs. 3b-c and 3e-f respectively. All polar plots show a distorted clover pattern reaching minima when the THz magnetic fields are aligned along the crystallographic $a$ and $c$ axes and maxima between the two axes. Notably, these maxima do not occur at the $ac$-bisector but are skewed towards the $a$ axis ($\theta \sim 30^\circ$) for the parallel configuration or $c$ axis for the perpendicular configuration ($\theta \sim 60^\circ$). This symmetry diverges from the polar patterns of other signals, as presented in Supplementary Fig. S4 and Fig. S5, which originate from the nonlinear excitations of individual magnon modes (see Supplementary Note 3 for details). In contrast, the polar patterns of SFG and DFG signals are a direct consequence of the fact that both second-order signals depend on two distinct magnon modes driven along orthogonal coordinates while only emitting along the $a$-axis. Combined, this yields azimuthal dependences, $A(\theta)$, of the form:
\begin{align*} 
	A_{\parallel}(\theta) &\propto \left|\chi_m^{(2)}H^2\sin\theta\cos^2\theta\right|, \text{ and } \\
	A_{\perp}(\theta) &\propto \left|\chi_m^{(2)}H^2\sin^2\theta\cos\theta\right|, 
\end{align*}
where $\chi_m^{(2)}$ is the corresponding second-order magnetic susceptibility (see Extended Data Fig. 2). 

These second-order nonlinear responses indicate the existence of coherent magnon-magnon mixing signals with a unique anisotropic response when both qFM and qAFM modes are driven out of equilibrium (see Fig. 4a). Figure 4b depicts the energy-level diagram of the nonlinear magnonic states with the presence of these coherent couplings. The linear interaction between the THz magnetic field and spins gives rise to a single-magnon one-quantum (1Q) coherence between the ground state and the excited magnon state, which radiates signals at frequency $\Omega_{qFM}$ or $\Omega_{qAFM}$, depending on which mode is being driven. This linear response dominates the single-pulse FID signals as shown in Fig. 1c-d. With the aid of the second field interaction, after the creation of the 1Q coherence between the ground state and the excited qFM (qAFM) mode, the second field can promote the SFG coherence involving the other magnon excitation, i.e., the qAFM (qFM) mode. In this case, the direct coupling between two magnon modes leads to the formation of a distinct correlated magnonic state, accumulating phase at their sum frequency $\Omega_{qFM}+\Omega_{qAFM}$ . Therefore, the appearance of SFG signals indicates a dynamical renormalization of coherent magnons, going beyond what can be achieved in thermal equilibrium. In addition, two magnon modes can be sequentially driven from the ground state without forming a new state and upon doing so they mutually interact, leading to the DFG coherence between the two magnon states that accumulates phase at the difference frequency $\Omega_{qAFM}-\Omega_{qFM}$. The emergence of these magnon-magnon mixing signals can be rationalized both classically and quantum mechanically, by taking into account the nonlinear interactions between different magnon modes. Both formalisms are provided in Supplementary Note 4.

It is worth emphasizing here that in centrosymmetric materials the even-order quantum coherences driven by electric dipole transitions do not radiate and are therefore not directly observable. In general, a third field interaction in the second time-delayed THz pulse is required to induce transitions from these even-order coherences into 1Q coherences that radiate the nonlinear (third order in the electric field) signals.\cite{turner2010coherent, stone2009two} Because the third field interferes coherently with the 2Q coherences, the amplitudes of the detected signals as a function of the inter-pulse delay time shows oscillations at the 2Q frequencies, allowing them to be measured indirectly. In contrast, the SFG and DFG signals observed here are coherent emissions from the second-order coherences. In YFeO$_3$, the lattice structure preserves inversion symmetry, but the canted spins result in net magnetization, allowing these second-order coherences to radiate. In this picture, only the THz magnetic field exerts an influence on the second-order signals. 

In order to better understand the underlying spin dynamics, we perform additional LLG simulations where we fix the THz-THz pulse delay at $\tau = 4.4$ ps, a point where the magnon-mixing signals are prominent. We then isolate the frequency components corresponding to the SFG and DFG signals for each sublattice spin and plot the resulting spin trajectories ($\mathbf{S}_1(t)$ in \textit{red} and $\mathbf{S}_2(t)$ in \textit{purple}) projected onto the $a$ and $c$ axes in Fig. 4c and 4d. The net magnetization dynamics, obtained by summing the contributions of the two spins, projected onto the same axes are shown in (\textit{blue}) for SFG and (\textit{purple}) for DFG. In these simulations, each spin's individual dynamics are highly elliptical. However, in all instances, their $c$ components cancel each other, which results in magnetization dynamics that occur purely along the $a$-axis, in agreement with the experimental anisotropic SFG and DFG emission patterns (see further discussions in Supplementary Note 5).

The current study highlights the strength of 2D THz spectroscopy in uncovering the coherent anharmonic couplings between distinct magnon modes. We envision that further investigations using our methodology can resolve other exotic dynamics of magnetic excitations, including magnon-phonon hybridization\cite{mai2021magnon,liu2021direct,cui2023chirality} and multiferroic soft modes\cite{kubacka2014large}, and provide insights into elusive states of matter, such as quantum spin liquids with long range entanglement\cite{wan2019resolving, choi2020theory, nandkishore2021spectroscopic, li2021photon}. Finally, the unique coherences observed here could result in future advances in quantum technology based upon nonlinear magnon mixing, such as all-magnonic mixers and converters.

\newpage
\noindent\textbf{Acknowledgments} 
Z.Z., Z.-J.L., M.T.W. and K.A.N acknowledge support from the U.S. Department of Energy, Office of Basic Energy Sciences, under Award No. DE-SC0019126. F.Y.G. and E.B. acknowledge support from the Robert A. Welch Foundation (grant F-2092-20220331). Y.-C.C. acknowledges direct funding from the MIT UROP. A.v.H. gratefully acknowledges funding by the Humboldt Foundation. Work by JC and PN was partially supported by the Department of Energy, Photonics at Thermodynamic Limits Energy Frontier Research Center, under Grant No. DE-SC0019140 and by the Quantum Science Center (QSC), a National Quantum Information Science Research Center of the U.S. Department of Energy (DOE). P.N. acknowledges support as a Moore Inventor Fellow through Grant No. GBMF8048 and gratefully acknowledges support from the Gordon and Betty Moore Foundation. T.K. acknowledges support from JSPS KAKENHI (21K14550, 20K22478).

\noindent\textbf{Author contributions} Z.Z. and F.Y.G. conceived the study and designed the research; Z.Z. and F.Y.G. performed the experiments and analyzed the data, assisted by Z.-J.L. and Y.-C.C.; Z.Z., F.Y.G., J.B.C., Y.-C.C., and M.T.W. performed theoretical analysis and simulated the LLG dynamics; T.K. and T.S. provided the sample; Z.Z., F.Y.G., J.B.C., A.v.H., P.N., E.B., and K.A.N. interpreted the data; Z.Z., F.Y.G., E.B. and K.A.N. wrote the manuscript; K.A.N. and E.B. supervised the project.

\noindent\textbf{Competing interests} The authors declare no competing interests.

\begin{figure}[hbtp]
	\centering
	\includegraphics[width=1\linewidth]{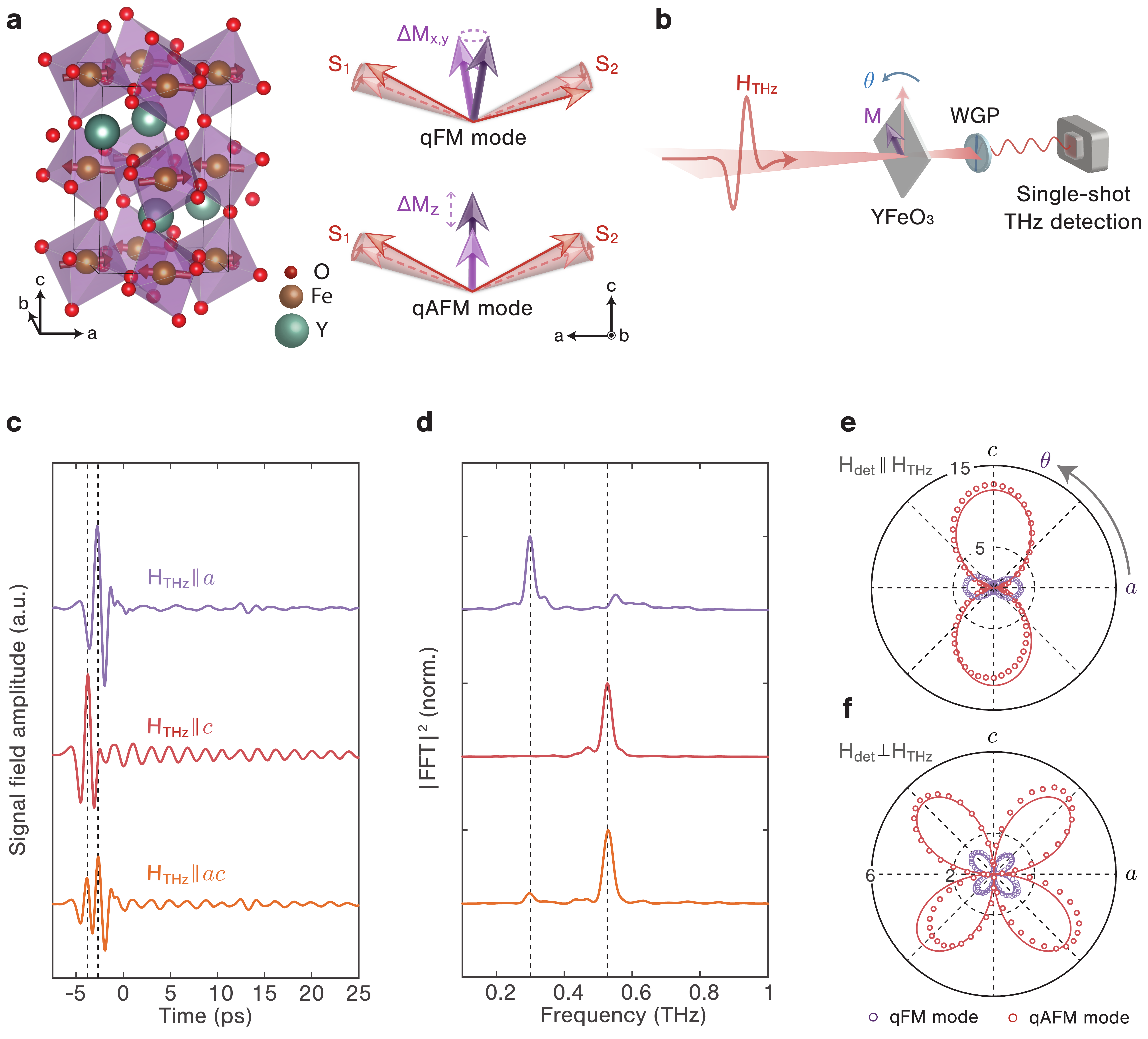} 
	\caption{\label{fig:Fig1}
	\textbf{Distinct magnon modes in the canted antiferromagnet YFeO$_3$. a,} Orthorhombically distorted perovskite structure of YFeO$_{3}$ (space group: $Pbnm$) with the net magnetization M along the crystal $c$ axis. Displacements along the qFM and qAFM magnon mode coordinates correspond to the out-of-phase or in-phase precession of the sublattice spins $\mathbf{S}_1$ and $\mathbf{S}_2$, resulting in an overall precession (qFM) or amplitude oscillation (qAFM) of \textbf{M}. \textbf{b,} Depiction of the experimental setup for the THz FID measurements. The orientation of the wire-grid polarizer (WGP) relative to the polarization of the incident THz field is used to select the parallel- ($\mathbf{H}_{det} \parallel \mathbf{H}_{THz}$) or perpendicular-polarized ($\mathbf{H}_{det} \perp \mathbf{H}_{THz}$) signal fields $\mathbf{H}_{det}$. \textbf{c,} The time-domain FID signals in the parallel-polarized configuration corresponding to the excitation of the qFM mode ($\mathbf{H}_{THz} \parallel a$ axis, \textit{purple}), qAFM mode ($\mathbf{H}_{THz} \parallel c$ axis, \textit{red}) and both ($\mathbf{H}_{THz} \parallel ac$ bisector direction, \textit{yellow}). The corresponding Fourier transforms (for signals after the THz pulses, i.e., after $t$ = 0) are in \textbf{d}, \textbf{e} and \textbf{f} show the azimuthal dependence of polarimetry with the parallel ($\mathbf{H}_{det} \parallel \mathbf{H}_{THz}$) and perpendicular ($\mathbf{H}_{def} \perp \mathbf{H}_{THz}$) geometries. The numbers in \textbf{e} and \textbf{f} are spectral amplitudes of the corresponding modes with an arbitrary unit.
}
\end{figure}

\begin{figure}[hbtp]
	\centering
	\includegraphics[width=1\linewidth]{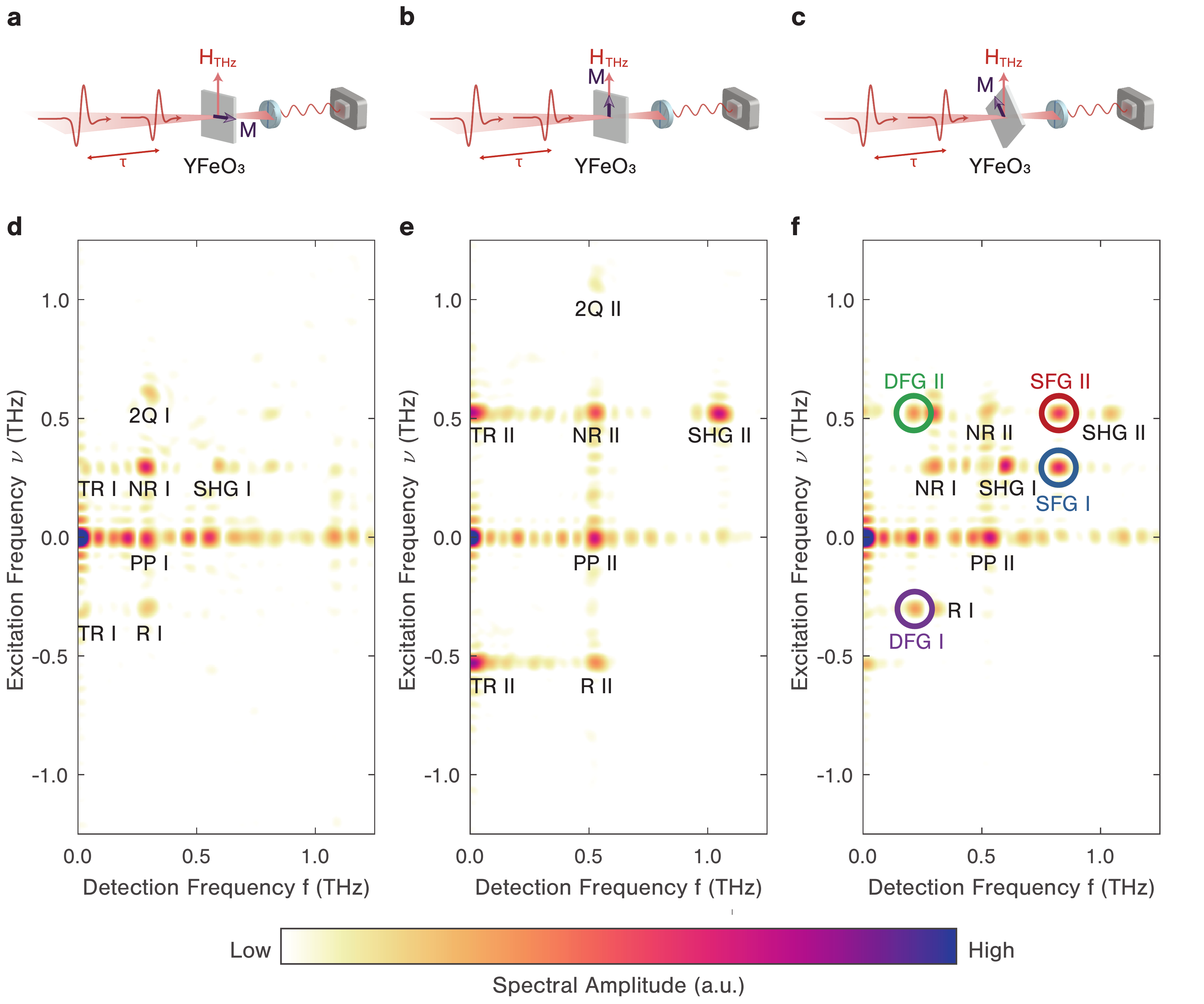} 
	\caption{\label{fig:Fig2}  
	\textbf{Nonlinear 2D THz spectra of YFeO$_3$. a-c,} Schematic illustrations of the excitation configurations for three different THz magnetic field orientations corresponding to $\mathbf{H}_{THz} \propto a$ axis, $\mathbf{H}_{THz} \propto c$ axis, and $\mathbf{H}_{THz} \propto ac$  bisector. The orange and purple arrows indicate the THz magnetic field orientation and the net magnetization, respectively. \textbf{d-f}, The corresponding nonlinear 2D THz spectra. Peaks corresponding to pump-probe (PP), rephasing photon-echo (R), non-rephasing (NR), two-quantum (2Q), second harmonic generation (SHG), and sum-frequency generation (SFG) and difference-frequency generation (DFG) signals are indicated. I and II refer to the qFM ($\Omega_{qFM}$) and qAFM ($\Omega_{qAFM}$) modes respectively. For the SFG and DFG signals, the assignment refers to the excitation frequency and indicates the time-ordering, i.e. which magnon mode was excited by the first THz field interaction. }
\end{figure}

\begin{figure}[hbtp]
	\centering
	\includegraphics[width=0.6\linewidth]{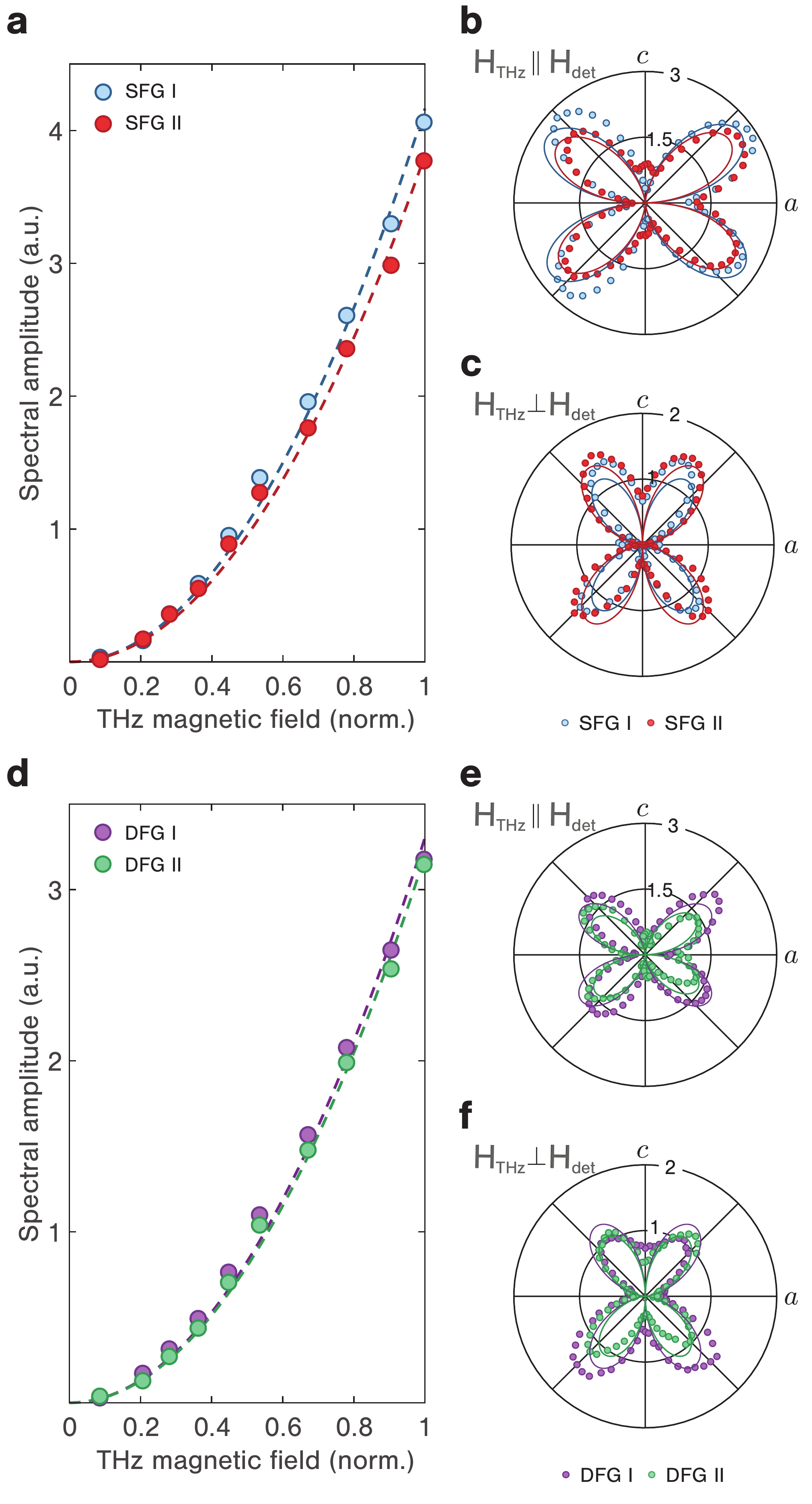} 
	\caption{\label{fig:Fig3} 
	\textbf{Field dependence and 2D THz polarimetry of the SFG and DFG signals.} The THz magnetic field dependences of the spectral amplitudes (\textit{circles}) of the \textbf{a,} SFG and \textbf{d,} DFG peaks obtained from 2D spectra with $\mathbf{H}_{THz} \propto ac$ bisector (i.e per Fig. \ref{fig:Fig2}c), are shown alongside quadratic fits of the form $\propto \mathbf{H}^2$ (\textit{dashed lines}). Polarimetry results showing the azimuthal dependences of the spectral amplitudes of the \textbf{b-c,} SFG and \textbf{e-f} DFG signals for both parallel and perpendicular-polarized analyzer configurations. Accompanying the experimental data (\textit{circles}) are fits to functions of the form $|\sin\theta\cos^2\theta|$ and $|\sin^2\theta\cos\theta|$ (\textit{solid lines}) for the respective polarization configuration. All spectral amplitudes were obtained directly from the nonlinear 2D THz spectra.  }
\end{figure}

\begin{figure}[hbtp]
	\centering
	\includegraphics[width=1\linewidth]{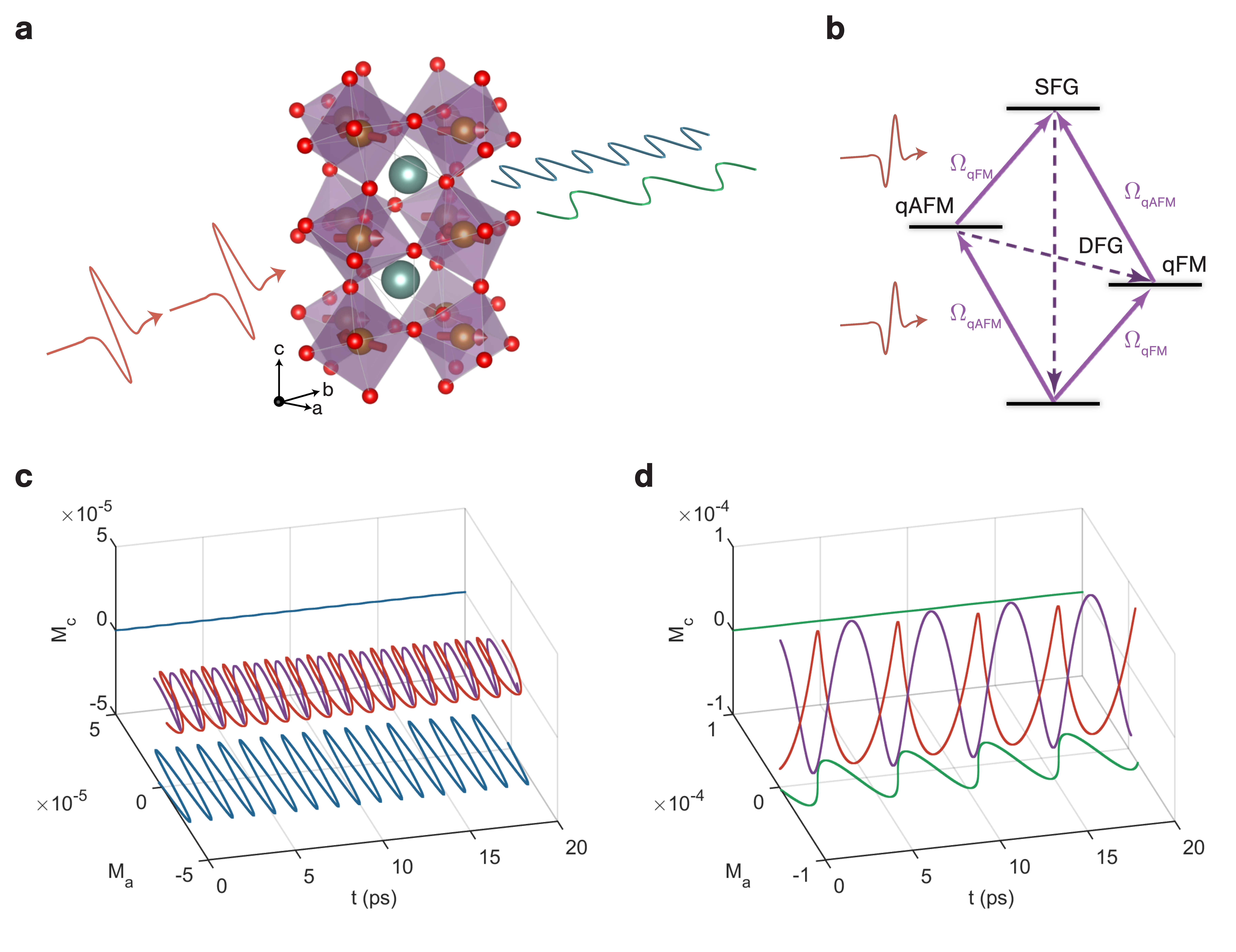} 
	\caption{\label{fig:Fig4}  
	\textbf{Excitation pathways of nonlinear magnonic states. a,} Excitation scheme with $\mathbf{H}_{{THz}} \parallel ac$-bisector showing the nonlinear SFG (\textit{blue}) and DFG (\textit{green}) emission with $\mathbf{H}_{{NL}} || a$-axis. \textbf{b}, Magnon energy level diagrams in the qFM and qAFM mode basis show the origins of different coupled magnon coherences. SFG signals originate from 2Q coherences which result from the nonlinear in-phase interference of stepwise qFM and qAFM excitations. Similarly, DFG signals originate from nonlinear out-of-phase interference of the same stepwise excitations. \textbf{c,} Time-dependent precession of sublattice spins $\mathbf{S}_1$ (\textit{red}) and $\mathbf{S}_2$ (\textit{purple}) projected along $a$ and $c$ are shown along with corresponding projections of the total magnetization (\textit{blue}) for the SFG signal. \textbf{d,} Per \textbf{c} but for the DFG signal with total magnetization shown in (\textit{green}). For both processes, nonlinear emission occurs only along the $a$-axis as the dynamics of the two sublattice spins cancel out for both $b$ and $c$.}
\end{figure}

\FloatBarrier
\captionsetup[figure]{name=Extended Data Fig.,labelfont=bf,labelsep=vline}
\newpage 
\section*{Extended Data Figures}
\FloatBarrier
\setcounter{figure}{0}
\begin{figure}[hbtp]
	\centering
	\includegraphics[width=1\linewidth]{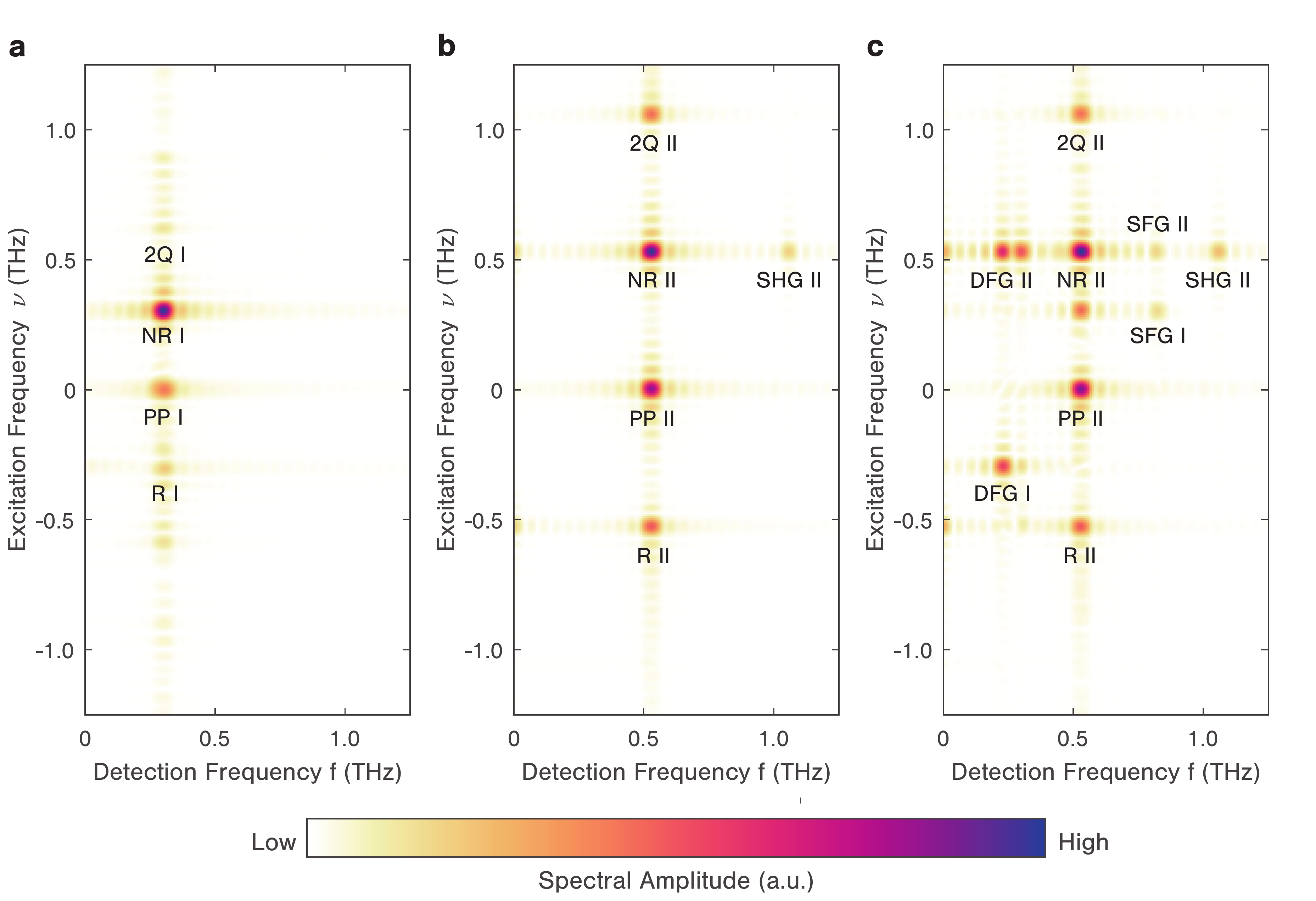}
	\caption{\label{fig:FigE1}  
	\textbf{Simulated 2D THz spectra of YFO.} Theoretical 2D THz spectra obtained from LLG simulations of $b$-cut YFO using the same THz magnetic field orientations shown in Fig. \ref{fig:Fig2}: \textbf{a,} $\mathbf{H}_{THz} \propto a$ axis, \textbf{b,} $\mathbf{H}_{THz} \propto c$ axis, and \textbf{c,} $\mathbf{H}_{THz} \propto ac$ bisector. Nonlinear mixing signals: SFG and DFG appear only upon simultaneous excitation of both qFM and qAFM modes when $\mathbf{H}_{THz} \propto ac$ bisector, in agreement with our experimental observations.
	}
\end{figure}

\begin{figure}[hbtp]
	\centering
	\includegraphics[width=1\linewidth]{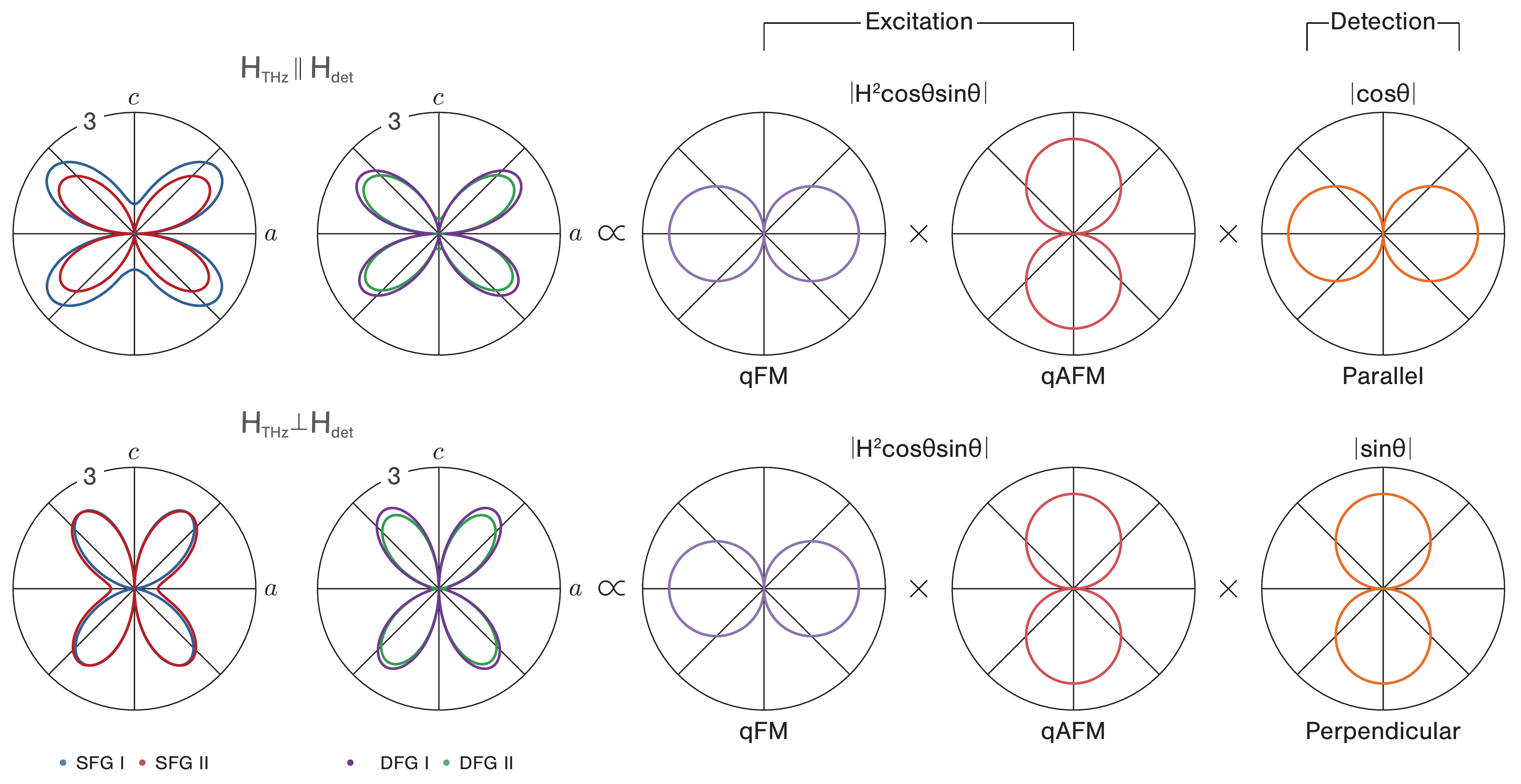}
	\caption{\label{fig:FigE2}  
		\textbf{Simulated polarimetry patterns.} (\textit{left}) Theoretical anisotropic SFG and DFG signal amplitudes are shown as a function of azimuthal angle $\theta$ for both parallel- ($\mathbf{H}_{THz} \parallel \mathbf{H}_\text{det}$) and cross-polarized ($\mathbf{H}_{THz} \perp \mathbf{H}_\text{det}$) detection configurations along with the (\textit{right}) corresponding decompositions into excitation and detection terms. For each $\chi^{(2)}_m$ magnon mixing signal, the generation mechanism is the result of the simultaneous excitation of qFM ($\cos\theta$) and qAFM ($\sin\theta$) modes, while nonlinear emission occurs along the crystallographic $a$-axis leading to separate symmetry terms for parallel- ($\cos\theta$) and cross-polarized ($\sin\theta$) signals.
	}
\end{figure}

\newpage
\normalsize

\section*{Methods}
\noindent \textit{Sample Preparation} \\
A single crystal of YFeO$_3$ (2-mm thick) grown by a floating zone melting technique was used in this work. The crystal was cut perpendicular to the $b$-axis. Before each THz measurement, the sample was magnetized to reinforce the residual magnetization and ensure the formation of a single magnetic domain.

\noindent \textit{Linear and 2D THz Polarimetry} \\
The laser source was a 1-kHz repetition rate Ti:Sapphire regenerative amplifier system which outputs 12-mJ, 35-fs pulses with a spectrum centered at 800 nm. The majority of the laser power was split evenly into two arms, which were then recombined with a controlled relative time delay on a single MgO:LiNbO$_3$ crystal in a tilted pulse-front geometry\cite{yeh2008generation} to generate a pair of time-delayed single-cycle THz pulses via optical rectification. The THz beams were then focused onto a sample with a pair of off-axis parabolic mirrors in a $4f$ configuration. The transmitted THz light was then recollimated and refocused by another pair of $4f$ parabolic mirrors onto a 2-mm thick ZnTe electro-optic (EO) sampling crystal where it was overlapped with the EO sampling probe beam derived from a small portion of the fundamental laser power. To obtain single-shot readout of the time-dependent THz signal waveforms, the optical probe beam was first expanded and then reflected off a 500-step staircase echelon mirror to generate 500 beamlets with pulses that were delayed successively by 40 fs, given by the step heights which determined the additional distances traveled by the successive pulses. The 500 probe pulses were then overlapped spatially with the THz beam in the EO crystal and relayed onto distinct regions of a CCD camera through a pair of $4f$ imaging systems. The THz field-induced birefringence in the EO crystal was measured by separating the beams into two orthogonal polarizations with a balanced detection scheme. Images of the probe beams were retrieved from the CCD camera at the 1-kHz repetition rate of the laser, permitting the full time-dependence of the THz signal within a 20-ps window to be measured on a shot-to-shot basis. The single-shot measurement method has been described previously.\cite{gao2022high}
For the linear THz measurements, one of the optical pulses that pumps the LN crystal is blocked and the signal generated through only the interaction with one THz pulse is measured. For 2D THz measurements, differential chopping of the two LN pump beams (A and B) is used to extract the nonlinear signal as the delay between the two THz pulses, $\tau$, is varied. The nonlinear signal $\mathbf{H}_{NL}$ is calculated via: 
\begin{equation}
	\mathbf{H}_{NL}(\tau,t) = \mathbf{H}_{AB}(\tau,t)  - \mathbf{H}_A(\tau,t)  - \mathbf{H}_B(t)  + \mathbf{H}_0(t),
\end{equation}
where $\mathbf{H}_{A}$, $\mathbf{H}_{B}$, $\mathbf{H}_{AB}$ and $\mathbf{H}_{0}$ are the THz signals measured with pump A, pump B, both pumps and no pumps, respectively. A 2D Fourier transform of the resulting time-domain signal yields the 2D THz spectrum. To avoid multireflection artifacts from the sample and EO sampling crystals, the collected THz signal was limited to a $\sim20$ ps time window in both $t$ and $\tau$, yielding an instrument linewidth of $\sim50$ GHz. This linewidth exceeds the natural linewidth of both magnon modes.  

For all polarimetry measurements, the sample was mounted on a rotational stage and 2D THz signals were measured as the crystal was rotated by 5$^\circ$ increments through a complete 360$^\circ$ revolution. A wire-grid polarizer placed after the sample acted as an analyzer to select the polarization of the THz emission detected, with the polarization of the optical probe beams and the orientation of the ZnTe crystal adjusted accordingly. The complete 2D THz polarimetry measurement included 144 2D spectra. Using the single-shot detection method, the total data acquisition time was about 22 hours. 
To measure the field dependence on the 2D spectra, a pair of wire grid polarizers was added before the sample to allow for the adjustment of the peak field level.

\noindent \textit{Numerical simulations} \\
A uniform two-sublattice Hamiltonian is considered here to simulate the nonlinear spin dynamics triggered by the magnetic field components of the THz pulses: 
\begin{align*}
		\mathcal{H} &= \mathcal{H}_0+\mathcal{H}_{Zeeman}\\
	 	&=nJ\mathbf{S}_{1}\cdot \mathbf{S}_{2}+n\mathbf{D}\cdot (\mathbf{S}_{1}\times \mathbf{S}_{2})-			\sum_{i=1,2}(K_{a}S_{ia}^2+K_{c}S_{ic}^2)\\
	 	&\hspace{12pt}-\gamma[\mathbf{H}_A(\tau,t)+\mathbf{H}_B(t)]\cdot(\mathbf{S}_{1}+\mathbf{S}_{2}),
\end{align*} 
where $\mathbf{S}_{i=1,2}$ represents each sublattice spin, $n$ is the number of nearest neighboring spin sites, $J$ is the exchange constant that stabilizes the antiferromagnetic order, $D$ is the antisymmetric exchange constant that accounts for the canting of the sublattice spins, $K_a$ and $K_c$ are magneto-crystalline anisotropies that determine the orientation of the net magnetization, $\gamma=\frac{g\mu_B}{\hbar}$ is the gyromagnetic ratio, and $\mathbf{H}_A$ and $\mathbf{H}_B$ represent the magnetic field components of two time-delayed THz pulses. Such a model has been shown to capture the essential physics of spin dynamics in canted antiferromagnets.\cite{kampfrath2011coherent, lu2017coherent, kurihara2018macroscopic} From this model Hamiltonian, the time-dependent effective magnetic field can be calculated as $\mathbf{H}_i^{eff}=-\frac{1}{\gamma}\frac{\partial\mathcal{H}}{\partial S_i}$, and thus the resulting Landau-Lifshitz-Gilbert (LLG) equation can be written as:
\begin{equation*} 
	\frac{d\mathbf{S}_i}{dt}=-\frac{\gamma}{1+\alpha^2}[\mathbf{S}_i \times \mathbf{H}_i^{eff}+\frac{\alpha}{|\mathbf{S}_i|}\mathbf{S}_i \times (\mathbf{S}_i \times \mathbf{H}_i^{eff})],
\end{equation*}
where $\alpha$ is a phenomenological Gilbert damping constant that accounts for energy dissipation.
To simulate the 2D spectra, we solve the above equations of motion for each sublattice spin $\mathbf{S}_i$ driven by a pair of THz pulses and then subtract the contribution from each THz pulse alone. Therefore, the nonlinear response of the net magnetization $\mathbf{M}=\mathbf{S}_1+\mathbf{S}_2$ can be extracted as a function of both $\tau$ and $t$. Performing 2D Fourier transforms with respect to $\tau$ and $t$ yields the simulated 2D THz spectra. 

To obtain the theoretical polarimetry patterns shown in Extended Data Fig. 2, we performed the above LLG simulations while rotating the crystal a full $360^\circ$ in 1$^{\circ}$ steps to obtain the amplitude of the nonlinear signal as a function of azimuthal angle $\theta$.

\noindent\textbf{Data availability}
Source data are provided with this paper. All other data that support the findings of this study are available from the corresponding authors on reasonable request.

\noindent\textbf{Code availability}
The codes used to perform the simulations and to analyse the data in this work are available from the corresponding authors upon request.

\newpage
\section*{References}
\footnotesize
\bibliographystyle{naturemag}
\bibliography{paper}

\end{document}